\documentclass[a4paper]{aa}
\usepackage{graphicx}
\usepackage{txfonts}
\usepackage{multirow}
\usepackage{url}
\newcommand{\mic}{~$\mu$m}
\newcommand{\dd}{^{\circ}}
\newcommand{\msol}{M$_{\odot}$}
\newcommand{\lsol}{L$_{\odot}$}

\newcommand{\kms}{km~s$^{-1}$}
\newcommand{\hii}{{\sc Hii}}
\newcommand{\GC}{Galactic Centre}
\newcommand{\GB}{Galactic Bulge}
\begin{document}

%\title{Infrared stellar population near the Galactic Centre
\title{Recent star formation in the inner Galactic Bulge seen
by ISOGAL
\thanks{This is paper no. 23 in a refereed
journal based on data from the ISOGAL project.}\fnmsep
\thanks{Based on observations with ISO, an ESA project with instruments
funded by ESA Member States (especially the PI countries: France,
Germany, the Netherlands and the United Kingdom) and with the
participation of ISAS and NASA.}
}
\subtitle{I - Classification of bright mid-IR sources in a test field}

\author{F. Schuller\inst{1,2} \and A. Omont\inst{1} \and I. S. Glass\inst{3}
\and M. Schultheis\inst{1,4} \and M. P. Egan\inst{5} \and S. D. Price\inst{6}}
\offprints{F. Schuller, schuller@iap.fr}
\institute{
Institut d'Astrophysique de Paris, CNRS and Paris-VI University,
98 bis, Bd Arago, F-75014 Paris, France
\and
Max Planck Institut f\"ur Radioastronomie, Auf dem H\"ugel 69,
DE-53121 Bonn, Germany
\and
South African Astronomical Observatory, PO Box 9, Observatory 7935, South
Africa
\and
Observatoire de Besan\c{c}on, CNRS, 41 bis, av. de l'Observatoire,
B.P. 1615, F-25010 Besan\c{c}on Cedex, France
\and
Air Force Research Laboratory, ODASD, 1851 s. Bell St., CM3,
Suite 7000, Arlington, VA 22202, USA
\and
Air Force Research Laboratory, Space Vehicles Directorate, 29 Randolph Road,
Hanscom AFB, MA 01731, USA
}

\abstract{The stellar populations in the central region of the Galaxy
are poorly known because of the high visual extinction and very
great source density in this direction.}
{To use recent infrared surveys for studying the dusty stellar objects
in this region.}
{We analyse the content of a $\sim20\times20$~arcmin$^2$ field centred at
$(l,b)=(-0.27\dd,-0.06\dd)$ observed at 7 and 15\mic ~as part of the
ISOGAL survey. These ISO observations are more than an order of magnitude
better in sensitivity and spatial resolution than the IRAS observations.
The sources are cross-associated with other catalogues to identify
various types of objects. We then derive criteria to distinguish
young objects from post-main sequence stars.}
{We find that a sample of about 50 young stellar objects and
ultra-compact \hii ~regions emerges, out of a population of evolved
AGB stars. We demonstrate that the sources colours and
spatial extents, as they appear in the ISOGAL catalogue, possibly
complemented with MSX photometry at 21\mic, can be used to determine
whether the ISOGAL sources brighter than 300~mJy at 15\mic ~(or
[15]$\leq$4.5~mag) are young objects or late-type evolved stars.}{}

\authorrunning{F. Schuller et al.}
\titlerunning{Recent star formation seen
by ISOGAL. I - FC-00027-00006}

\maketitle

\section{Introduction}
\label{sint}

The ISOGAL survey is a large set of observations of the inner Galaxy at
7 and 15\mic ~(Omont et al. \cite{PSC_AO}, Schuller et al. \cite{PSC_FS}),
obtained with the ISOCAM instrument on board the ISO satellite
(Cesarsky et al. \cite{ISOCAM}).
The ISOGAL fields are distributed along the Galactic plane,
mostly in the $\vert b \vert \la 1\dd$ range.
The achieved sensitivity is about 10~mJy in most fields in the Galactic
Disk, and of order 20--30~mJy in the fields observed with narrow filters
and 3$''$ pixels, including the \GC ~region, and thus the observations
discussed in the present paper.
The catalogues of more than $10^5$
extracted point sources have been systematically
cross-identified with the near infrared DENIS data
(Epchtein et al. \cite{ref-denis2}),
resulting in the ISOGAL--DENIS Point Source Catalogue (PSC),
which provides measurements at 7 and/or 15\mic, together with
up to three near infrared bands ($I$, $J$ and $K_{\rm s}$).
The astrometric accuracy is better than 0.5$''$ for sources
with a DENIS counterpart, and of order 2$''$ for sources without.

\begin{figure*}[!htbp]
\begin{center}
\resizebox{18cm}{!}{ \rotatebox{0}{\includegraphics{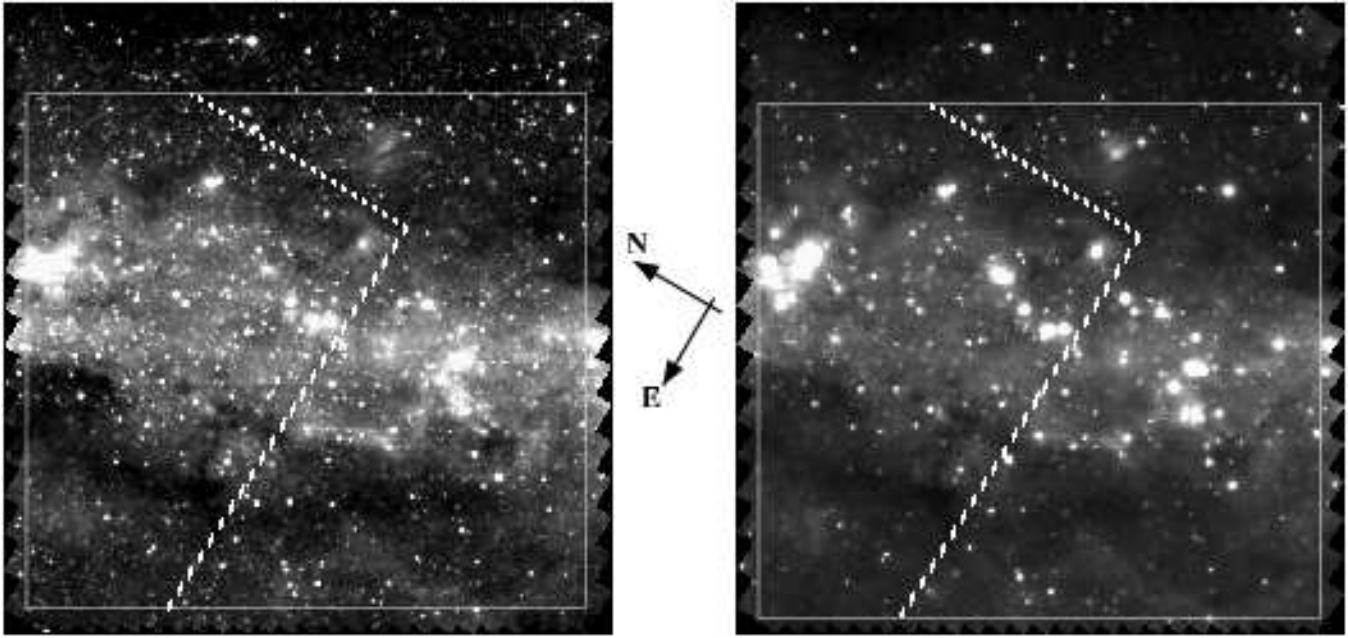}}}
\caption{\label{images}ISOCAM raster images used for the FC--027 field:
LW5 (7\mic, left) and LW9 (15\mic, right). The rectangular frames
show the limits of the ``regular'' field, as defined in Schuller et al.
(\cite{PSC_FS}) - see also Eq. (\ref{eq-limits}).
The dotted lines show the limits of the area included in the Glass et al.
(\cite{GLASS2}) survey (see Sect. \ref{sec-lpv}): only the area to the left
of these lines was observed by Glass et al. (\cite{GLASS2}).
% The upper parts of these images outside the rectangular frames overlap
% another field and are not considered here.
North and east directions are indicated.}
\end{center}
\end{figure*}

The analysis of various individual ISOGAL fields in the Galactic disk
and bulge have shown that the bulk of the detected
stellar population is composed of AGB and late RGB stars
(see references in Omont et al.~\cite{PSC_AO}).
They show a rather well defined sequence in the [15] vs. [7]--[15]
colour magnitude diagram, in good agreement with the
colours expected for such objects (Fig.~\ref{cmd_1}).
On the other hand, a few thousand ISOGAL sources can be interpreted
as young stellar objects or young massive stars (hereafter young
objects) deeply embedded in dusty envelopes or thick disks.
Theoretical models, as well as a systematic analysis of ISOGAL images
and catalogues, enabled us (Felli et al. \cite{FELLI2000},
\cite{Felli2002}; Schuller \cite{ref-myPhD})
to define criteria to roughly distinguish candidate young
objects from the bulk of
post-main sequence stars in the [15] vs. [7]--[15] diagram.

The ISOGAL FC--00027--00006 field (hereafter called FC--027), centred at
($l$,$b$) = $(-0.27\dd,-0.06\dd)$, is located very close to the \GC,
between Sgr A$^*$ and Sgr C, in a region of strong diffuse emission
at infrared and radio wavelengths. The visual extinction in this field
ranges from 20 to more than 35 magnitudes (Schultheis et al.
\cite{SCHULTHE}), and the photometry must be corrected carefully 
in order to interpret the observed magnitudes and colours properly,
especially in the near infrared. However, the [7]--[15] colours cannot
be affected by more than $\sim$1~mag of {\em interstellar} extinction
(see Fig.~\ref{cmd_1}).

We will show in the present paper that, in spite of the extinction, the
nature of the bright sources (with magnitudes at 15\mic ~[15]~$\leq$~4.5,
equivalent to flux densities above 280~mJy)
can be inferred from the raw data, as they appear
in the ISOGAL PSC. The bulk of the brightest sources can be
identified with various classes of objects (long period variable stars,
OH/IR stars, ultra-compact \hii ~regions). The mid-infrared properties
(especially the [7]--[15] colour and spatial extension at 15\mic)
of young objects (YSOs and \hii~regions) are quite different
from those observed for late-type stars. This leads to robust
selection criteria for extracting the young population from the ISOGAL
PSC (see also Lumsden et al.~\cite{ref-lumsden} for a similar
analysis based on the MSX data).
These criteria will be used to make a
census of such objects in all ISOGAL fields covering the
inner \GB ~in a forthcoming paper, where we will estimate
their luminosities, taking into account extinction corrections.

The ISOGAL data are briefly presented in Sect.~\ref{sec-iso-obs}.
We then show in Sect.~\ref{sec-known-src} that most ISOGAL bright 
sources can be associated with known objects, in particular long
period variable stars, radio-sources, or infrared sources from the
IRAS and MSX point source catalogues.
Our main results, consisting of criteria to select
young sources from the ISOGAL catalogue, are summarised
and used to extract luminous young candidate objects
without known counterparts from the whole FC--027
field in Sect.~\ref{sec-summary2}.

\section{\label{sec-iso-obs}ISOGAL observations}

A complete description of the ISOGAL observations, their processing and
the creation of the Point Source Catalogue (PSC) can be found in
Schuller et al. (\cite{PSC_FS}). In the present paper, we analyse
the source content of the FC--027 field.
We confine our analysis to the formal limits of this field, or the
``regular'' catalogue (as opposed to the ``edge'' catalogue, see
Schuller et al.~\cite{PSC_FS}),
which covers 18.5$\times$17 arcmin$^2$ (0.09~deg$^2$)
centred at $(l,b) \, = \, (-0.27\dd,-0.06\dd)$.
Namely, the present paper deals with the area defined by:
\begin{equation}
\label{eq-limits}
\left\{  \begin{array}{l}
-0.424 \; \leq \; l \; \leq \; -0.115 \\
-0.194 \; \leq \; b \; \leq \; 0.089
\end{array} \right.
\end{equation}

This field was observed with the narrow filters
LW5 (6.5-7.1\mic) and LW9 (13.9-15.9\mic), using pixels of 3$''$
field of view (see the raster images in Fig.~\ref{images}).
The 7\mic ~image was observed on 1996, Sept. 24, while the 15\mic
~observation was done on 1998, Feb. 19, so that variability
can affect the observed colours.
Point sources were extracted from the 7\mic ~and 15\mic ~observations
down
to magnitudes 8.4 at 7\mic ~(35~mJy), and 7.0 at 15\mic ~(28~mJy),
corresponding to roughly 50\% completeness (Schuller et al.
\cite{PSC_FS}).
About 90\% of the sources also have DENIS (Epchtein et al.
\cite{ref-denis2}) near-infrared counterparts, with
$K_{\rm s}$ magnitudes brighter than 11.3. Fainter DENIS sources
were not considered in order to reduce the fraction of chance
associations, so that a few faint near-infrared counterparts
may have been missed. Altogether, we estimated that no more
than $\sim$1\% of ISO--DENIS associations should be wrong.

% chance proba to find a DENIS source within 3.6'' = pi*(3.6/3600)^2*72000 = 22.6%
% 9.8% of ISO sources have no DENIS => among the 1843 with associations,
% 9.8%*22.6% = 2.21% should be false (= 40 assoc.)

%The search radius used 
%for the ISO-DENIS associations was 3.6'' and resulted in a probability
%of chance associations around 20\%, but dedicated quality flags were
%computed in order to greatly reduce this probability when considering
%only associations with good quality flags (Schuller et al. \cite{PSC_FS}).

%%%
% \section{Asymptotic Giant Branch stars}
%%%
\section{\label{sec-known-src}Identification of known sources}

In this Section, we will show how ISOGAL point sources can be associated
with sources from various catalogues, taken from the abundant
literature dealing with the centre of the Galaxy. In particular, we will
focus on: long period variable stars (LPVs) and
OH/IR stars (Sect.~\ref{sec-lpv}),
radio sources detected in the cm continuum
(Sect.~\ref{sec-radio}), the IRAS Point Source Catalogue
(Sect.~\ref{sec-iras}) and the MSX Point Source Catalogue
(Sect.~\ref{sec-msx}).
Additional results derived from specific near infrared
spectroscopic observations are discussed in Sect.~\ref{sec-sofi}.
The main results outlined from this
compilation of data are summarised in Sect.~\ref{sec-summary1}.

%\subsection{LPVs}
%\subsection{\label{sec-lpv}Long Period Variables}
\subsection{\label{sec-lpv}Late evolved stars}

As in the other ISOGAL fields, the huge majority of the detected point
sources in the FC--027 field are luminous late-type giants toward the
end of the red giant phase, mostly in the AGB phase
(Omont et al. \cite{OMONT1},
Glass et al. \cite{GLASS}, Ojha et al. \cite{ref-ojha}, van Loon et al.
\cite{ref-JvL}). Their photospheres are bright emitters in the near-infrared,
and many of them are surrounded by dusty circumstellar envelopes
due to mass loss.
The dust grains absorb the stellar radiation
and re-emit in the mid-infrared, so that these stars are bright in
mid-infrared images. Most of these sources are variable (Alard et
al.~\cite{ref-alard}), and we have been
able to find ISOGAL counterparts to variable stars of different classes
(Miras, OH/IR, other long period variables).

%\subsubsection{\label{lpv-assoc}Associations with ISOGAL sources}

Glass et al. (\cite{GLASS2}) have monitored an area of 24x24 arcmin$^2$
centred at the Galactic Centre over four years, in the K band, in search of
variable stars. Their surveyed area overlaps the FC--027 field by
$\sim$0.042~deg$^2$, or nearly one half the ISOGAL field (see
Fig.~\ref{images}). In this overlap area, they found 110 long period variable
stars (hereafter LPVs).
% 16 of which are also OH/IR stars.

\begin{figure}[htbp]
\begin{center}
\resizebox{8.5cm}{!}{ \rotatebox{0}{\includegraphics{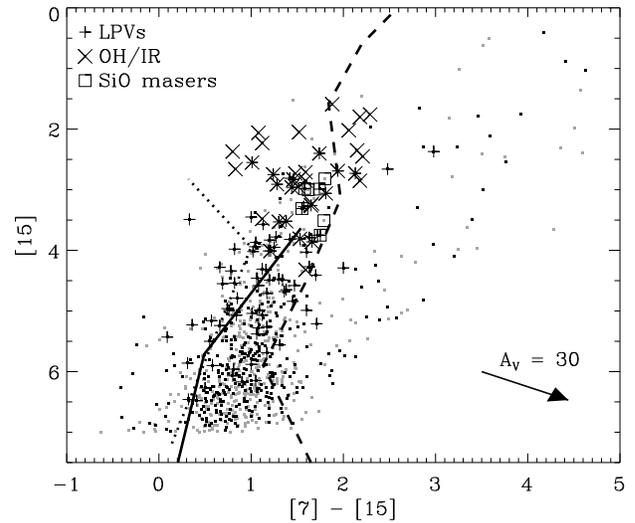}}}
\caption[]{\label{cmd_1}
[15] vs. [7]--[15] colour magnitude diagram for the FC--027 field.
Sources located inside the area also observed by Glass et al.
(\cite{GLASS2}) are shown with black symbols, while those outside
the Glass et al. (\cite{GLASS2})area are indicated with gray
symbols. The sources associated
with LPVs are indicated by plus signs, OH/IR stars are shown
with crosses, and sources where an 86~GHz SiO maser was detected
are shown with square symbols.
The thick lines show the loci of giant stars (continuous line),
supergiants (dotted line), M-type AGB stars and L-type stars with
different mass-losses (dashed line) in this
diagram, as computed using synthetic spectra (M. Cohen, private
communication). The extinction vector in the bottom right corner
corresponds to A$_{\rm V}$~=~30~mag; it was computed using the
mean values of A$_\lambda$/A$_{\rm V}$ for the LW5 and LW9 filters
derived by Jiang et al.~(\cite{jiang05}).}
\end{center}
\end{figure}

Among these 110 LPVs, 97 are associated with ISOGAL sources within a 5$''$
search radius, 89 of which are detected at both 7 and 15\mic.
These have [7]--[15] colours mostly around 1~mag, along a
sequence spanning the range 3 to 6~mag at 15\mic ~(Fig.~\ref{cmd_1}).
However, a few 0.1~mag in the observed colours may be due to
variability, because of
the 1.5 year delay between the 7\mic ~and 15\mic ~observations.
Blending effects may also introduce errors in the [7]--[15] colours.
Most of the 13 LPVs without association appear as faint objects at 7\mic.
Confusion with the bright diffuse background, added to the high source
density in this field, explain these peculiar cases. In addition,
10 out of these 13 sources have mean K-band magnitudes above 9.5
(Glass et al.~\cite{GLASS2}), while only one third of those with
an ISOGAL counterpart have such faint near-IR magnitudes.

%\subsection{\label{sec-ohir}OH/IR stars}

Three catalogues of OH/IR stars (Lindqvist et al. \cite{lindqvist},
Sevenster et al. \cite{sevenster}, Sjouwerman et al. \cite{sjouw})
were compiled and associated with ISOGAL data in the inner Galactic
Bulge by Ortiz et al. (\cite{ortiz}).
They found 37 OH/IR stars in the FC--027 field,
34 of which are associated with ISOGAL sources within 6$''$.
As shown with crosses in Fig.~\ref{cmd_1}, they
are found at the brightest end of the main sequence in this diagram,
with magnitudes [15] $\leq$ 4.5 and
colours 0.8 $\leq$ [7]--[15] $\leq$ 2.3. 
These are typical colours and magnitudes for OH/IR stars with bolometric
magnitudes in the range $-5.0 \pm 1.0$~mag and mass-loss
rates between $\sim 10^{-6}$ and a few $10^{-5}$~\msol/yr
(Ortiz et al. \cite{ortiz}, Groenewegen \cite{groen}).

%\subsection{\label{sec-sio}SiO masers}

Another seven sources (indicated with square symbols in Fig.~\ref{cmd_1})
were detected in the 86~GHz SiO maser line by Messineo et al.
(\cite{ref-messineo}), who observed 12 sources selected in the FC--027
field. This detection rate of 58\%
is consistent with the 48\% detection rate found by
Deguchi et al. (\cite{ref-deguchi}), who surveyed the complete
sample of LPV stars published by Glass et al. (\cite{GLASS2})
in SiO 43~GHz lines.

\begin{table*}[htbp]
\caption{\label{tab-radio-sup}Associations between uncatalogued
radio sources
visible in the 1616~MHz map of Liszt and Spiker (\cite{LS95}) and bright
ISOGAL sources. The 1950 coordinates (Cols.~1 and 2) are extracted
from Fig. 1 in Liszt and Spiker (\cite{LS95}), and were precessed to
2000 epoch in Cols.~3 and 4.}
\begin{center}
\begin{tabular}{llll@{ }c@{ }lll}
\hline
\hline
$\alpha_{1950}$ & $\delta_{1950}$ & $\alpha_{2000}$ & $\delta_{2000}$ 
& & ISOGAL Name & [7] & [15] \\
\hline
17$^h$41$^m$54$^s$ & --29$\dd$09.0$^\prime$ & 17$^h$45$^m$05$^s$
& --29$\dd$10$^\prime$12$^{\prime\prime}$ & & 174505.6-291018 &
5.24 & 1.78 \\
17 41 43 & --29 10.5 & 17 44 54 & --29 11 43 & \multirow{2}*[3pt]
{\huge{$\left. \right\}$}} & \multirow{2}*{174452.5-291122} &
\multirow{2}*{5.70} & \multirow{2}*{2.11} \\
17 41 38 & --29 10.0 & 17 44 49 & --29 11 13 & &    &      &      \\
17 41 23 & --29 12.6 & 17 44 34 & --29 13 50 & &
174433.8-291355 & 6.49 & 1.98 \\
\hline
\end{tabular}
\end{center}
\end{table*}

%%%
% \section{Young Stellar Objects}
%%%
\subsection{\label{sec-radio}Radio continuum sources}

Large fractions of the Galactic Disk, including the central region, have
been surveyed at radio wavelengths by different teams, using the VLA
at 1.4~GHz (Zoonematkermani et al. \cite{zoone}), 5~GHz (Becker et al.
\cite{BWHZ94}), and 1.6~GHz (Liszt \cite{Liszt85}; Liszt \& Spiker
\cite{LS95}). In addition, Downes et al.~(\cite{DGSW}) conducted
pointed observations in the \GC ~region with several telescopes.

A number of compact or small extended radio continuum sources
from these catalogues are found in the FC--027 field.
We consider as their infrared counterparts
the nearest ISOGAL neighbour brighter than 5~mag at 15\mic.
Infrared counterparts are found within 10$''$ for all sources,
except for source number 3 in Downes et al. (\cite{DGSW}) for
which the separation is 18$''$, and source number 7 in
Liszt \& Spiker (\cite{LS95}) which coincides with a compact
group of ISOGAL sources discussed in Sect.~\ref{sec-cluster}.

\begin{figure}[htbp]
\begin{center}
\resizebox{8.5cm}{!}{ \rotatebox{0}{\includegraphics{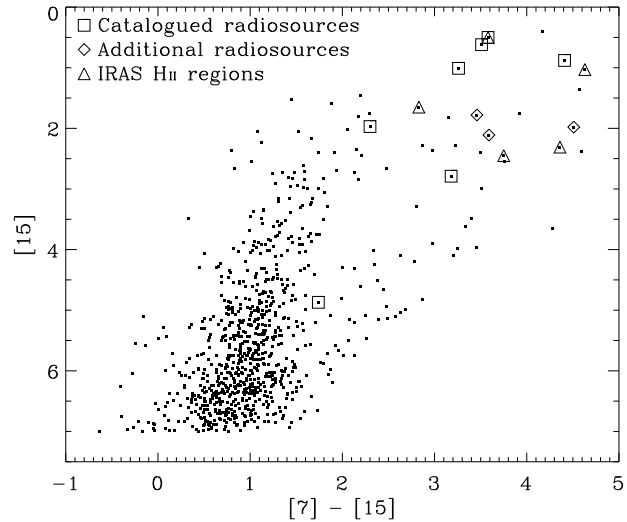}}}
\caption[]{\label{cmd_yso}
[15] vs. [7]--[15] colour magnitude diagram for all sources with
7\mic ~and 15\mic ~detections and reliable 7-15\mic ~associations
in the FC--027 field. The sources associated with radio sources
or with \hii ~regions detected by IRAS are indicated with various
symbols, as indicated in the top left corner.}
\end{center}
\end{figure}

In addition to these catalogued sources, a number of faint radio sources
can be seen in the 1616~MHz map of Liszt and Spiker (\cite{LS95}, see
their Figure 1). They were not included in their table, but some of them
can be associated with bright ISOGAL sources, as indicated in
Table~\ref{tab-radio-sup}. The positions of all these radio sources in the
[15] vs. [7]--[15] colour magnitude diagram are shown in Fig.~\ref{cmd_yso}.

The sensitivity of the Becker et al. (\cite{BWHZ94}) catalogue
in the central
region of the Galaxy is of order 40~mJy, and that of the Downes et al.
(\cite{DGSW}) survey is close to 10~mJy. Liszt \& Spiker (\cite{LS95})
report a noise level of 0.7~mJy/beam in their data, but their
observations cover only a fraction of the FC--027 field. In addition,
all these radio data show extended emission at various positions
in the field, so that confusion may limit the actual sensitivity
of these surveys. It is clear that any \hii ~region 
powered by a single B-type star at the distance of the \GC ~would
have been missed by these radio observations.

\begin{figure}[htbp]
\begin{center}
\resizebox{8.5cm}{!}{\includegraphics{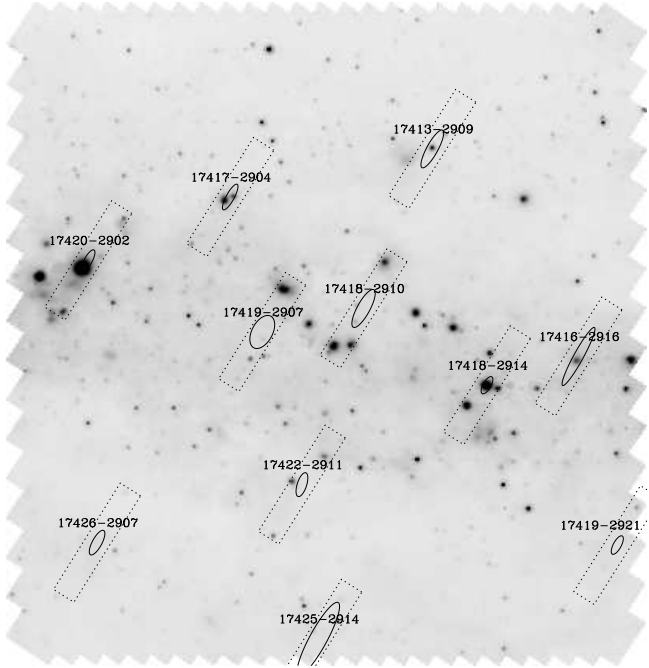}}
\caption{\label{fig-iras-ellips}Positions of IRAS point sources
found in the FC--027 field. The gray scale background shows the
15\mic ~ISO image. Every IRAS source is shown with its name, and
the 3$\sigma$ uncertainty ellipse. The IRAS field of view at
12 and 25\mic ~(0.76$\times$4.55~arcmin$^2$) is also shown
with dotted lines.}
\end{center}
\end{figure}

\subsection{\label{sec-iras}IRAS sources}

The IRAS observations were highly limited by source confusion
in the very dense regions of the inner Galactic Bulge, so that no
clear and unique association can be found between the IRAS and
ISOGAL catalogues in most cases. We therefore looked for ISOGAL
counterparts at least within the 3$\sigma$ uncertainty
ellipse of each IRAS point source.
In most cases,
both the positions and the photometry of the IRAS sources
probably result from a combination of several infrared sources,
on top of diffuse emission (Fig.~\ref{fig-iras-ellips}).
We will thus consider the brightest source at 15\mic ~as the most
likely main counterpart to the source responsible for the emission
detected by IRAS.

Eleven IRAS PSC sources are located in this field.
Four of them (17418--2914, 17419--2907, 17420--2902 and 17422--2911)
have $F_{25}/F_{12}$ and $F_{60}/F_{25}$ flux ratios
all higher than or close to 10, making these sources very likely
to be associated with compact \hii ~regions
(Wood and Churchwell \cite{ref-wood-church}).
Two other sources (17416--2916 and 17417-2904), with $F_{25}/F_{12}
> 3$ and no detection at 60\mic, could well be compact \hii ~regions.
The ISOGAL counterparts to these ultra-compact \hii ~regions
show peculiar
colours: they are very bright at 15\mic, as was expected from their
detection by IRAS, and very red in [7]--[15] (see triangle
symbols in Fig.~\ref{cmd_yso}).

\begin{figure}[htbp]
\begin{center}
\resizebox{8cm}{!}{ \rotatebox{0}{\includegraphics{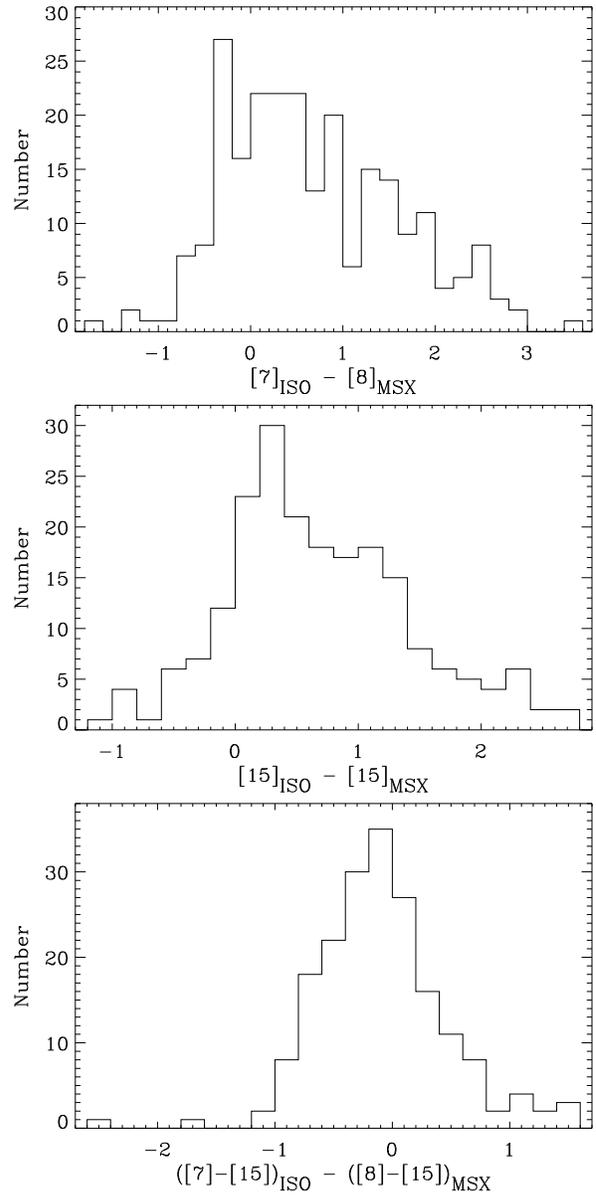}}}
\caption{\label{fig-mag-iso-msx-deep}Distributions of differences in
magnitudes between ISOGAL and MSX: ISO 7\mic--MSX A-band
(top), ISO 15\mic--MSX D-band (middle), and ISO 7--15 colour--MSX
A--D colour (bottom).}
\end{center}
\end{figure}

\subsection{\label{sec-msx}MSX sources}

The Midcourse Space Experiment (Price et al. \cite{ref-price})
performed observations in the mid-infrared
of the complete Galactic plane, in the $\vert \, b \, \vert \, \leq 5\dd$
range, in six bands centred on wavelengths between 4 and 21\mic,
at a spatial resolution of $\sim 18''$.
Here we will use version 2.3 of the MSX PSC
(Egan et al.~\cite{ref-msx2}). It
was built by extracting point sources from co-added
images, resulting in a 400~000 source catalogue with
a sensitivity of about 40~mJy in the A band (8.3\mic).
In particular, three raster-scanned experiments were combined at
the \GC, greatly improving the sensitivity.
%%% \subsubsection{MSX catalogue version 2.3}
%%%%%%
However, as many as 10\% of the point sources extracted from the densest
region of the \GC ~may actually be diffuse emission on the scale of the
MSX pixel rather than true point sources, as we will show below.
%%%%%%

We find 333 MSX sources in the FC--027 field.
After adding small offsets $(\Delta \alpha , \Delta \delta) 
= (+1.44'',$ $+0.32'')$
to the MSX coordinates, 257 sources (77\%) can be associated with
ISOGAL sources within an 8$''$ search radius. Two among the
remaining 76 are located near the edge of the field and
can be associated with ISO sources outside the limits.
A few other ones, including the two brightest ones at 15\mic
~among the non-associated MSX sources, are found at the barycentre of
two or more bright sources that appear resolved with ISO. This is
clearly an effect of the higher spatial resolution of ISO as compared
to the 18$''$ pixels of MSX.

However, at least one ISOGAL point source can always be found within
30$''$ of any MSX source. We have performed a $\chi^2$-analysis,
similar to that used to cross-corelate the MSX and 2MASS catalogues
in the Large Magellanic Cloud (Egan et al.~\cite{ref-chi2}),
to estimate the probability that an MSX source and its closest
ISOGAL neighbour can be real matches. It turned out that,
among the 76 sources with no counterpart at less than 8$''$,
55 associations have $\chi^2$ values below 18.6, where 99.99\%
of the true matches must lie. All of them have ISO--MSX
separations less than 18$''$, i.e. are within one MSX pixel,
so that these associations may be real, except for about ten
of them that show ISO--MSX magnitude differences greater
than 2~mag. Altogether, about one half of the latter associations
with separations greater than $8''$ may still be real.
Most of the 76 MSX sources without ISOGAL associations
within 8$''$ are located in regions of diffuse
emission, and the other half may well correspond to patches
of diffuse emission rather than genuine point sources.

To estimate the fraction of chance associations between MSX and
ISOGAL, we also performed cross-identifications between the ISOGAL PSC
and the MSX catalogue, adding an arbitrary offset to the coordinates
of the latter. The results show that nearly one third of the MSX sources
can be randomly associated with ISO sources within an 8$''$
search radius, as expected from the average ISO source density.
However, when taking into account the fact
that a fraction of these sources are well associated with
ISOGAL, we estimate that no more than 40 ISOGAL--MSX associations
(out of 257) should be false. Also note that the number of
MSX sources found within
the FC--027 field corresponds to {0.1 source/pixel in average,}
so that the extraction was certainly limited by confusion.
Therefore, the reality of the weakest sources and their associations with
ISOGAL should be regarded with extreme caution.

Nevertheless,
we find a fair general agreement between the ISO and MSX magnitudes,
both at 7 and at 15\mic, as shown in Fig.~\ref{fig-mag-iso-msx-deep},
although with quite large dispersion partly due to the variability of some
sources, and to a few wrong associations for the faintest sources.
The distributions of magnitude differences show clear
tails toward positive differences, that can be explained by
MSX including more flux from diffuse extended emission, or
from several point sources resolved by ISO, while only the magnitudes
of the nearest ISO counterparts have been taken into account.
In addition, the false ISO--MSX associations, as well as the
Malmquist bias on the faintest MSX sources, certainly contribute
to the tail in that direction (MSX magnitudes seem brighter
than ISOGAL), due to the lower sensitivity of MSX compared
to ISOGAL.

\begin{figure}[htbp]
\begin{center}
\resizebox{8.5cm}{!}{ \rotatebox{0}{\includegraphics{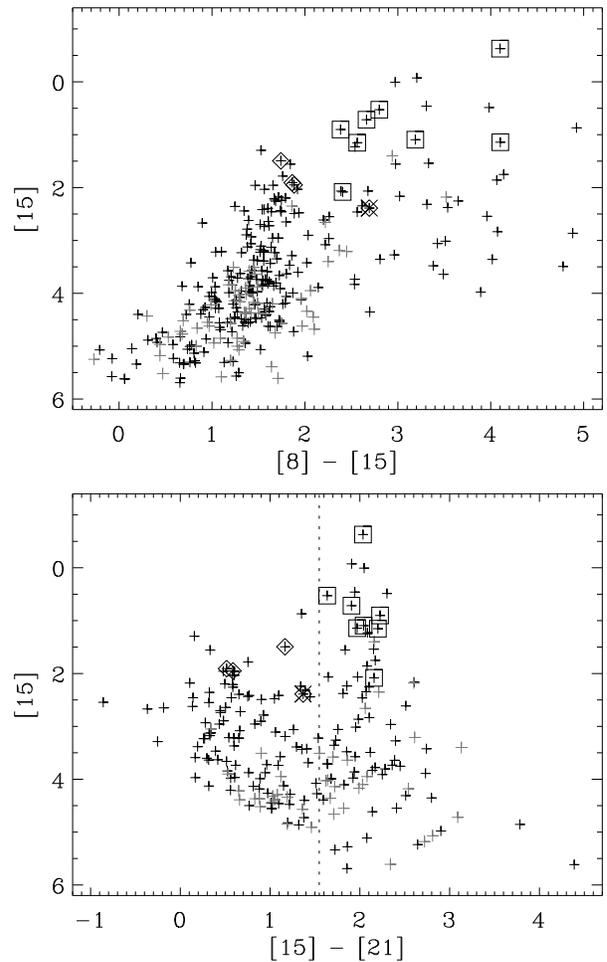}}}
\caption{\label{cmd-msx-deep}MSX colour-magnitude diagrams of the
FC--027 field: [15] vs. [8]--[15] (top) and [15] vs. [15]--[21]
(bottom). Sources identified with known \hii ~regions,
as derived from IRAS or radio continuum measurements, are shown
with square symbols, and those associated with LPV or OH/IR stars are
shown with $\times$ symbols or diamonds, respectively.
The grey crosses show the positions
of MSX sources without ISOGAL counterparts within 8$''$.
The dotted line in the
bottom panel corresponds to a flux ratio F$_E$/F$_D$~=~2 (see text).}
\end{center}
\end{figure}

The MSX survey provides photometry up to 21\mic ~(E band, 18.2--25.1\mic),
which can be used to extract candidate young massive stars
with a colour criterion
similar to the Wood and Churchwell (\cite{ref-wood-church}) criteria
for IRAS colours. Indeed,
the ISOGAL sources identified as young objects from their IRAS or radio
counterparts all have MSX flux ratios F$_E$/F$_D$ above 2 (or
colours [15]--[21]~$>$~1.55~mag, see
Fig.~\ref{cmd-msx-deep}). Therefore we will
use the criterion F$_E$/F$_D$~$\geq$~2 as an indication of
the young nature of
ISOGAL-MSX sources in Sect.~\ref{sec-summary2}.
The interpretation as young objects from the MSX D-E colour alone is somewhat
less secure than the Wood and Churchwell (\cite{ref-wood-church})
criteria, since objects of this type emit the bulk of their
energy at wavelengths longer than 25\mic.
Also, not all sources with F$_E$/F$_D$~$\geq$~2 show other
signs of high mass star formation, and the nature of those
with moderate [8]--[15] colours ($\la$~2~mag) remains unclear (see also
Sect.~\ref{sec-extract-all} and Fig.~\ref{fig-ccd-msx}).
However, the limit F$_E$/F$_D$~$\geq$~2 is more conservative than
the criteria F$_E$~$>$~F$_D$~$>$~F$_A$ and F$_E$~$>$~2$\times$F$_A$
used by Lumsden et al. (\cite{ref-lumsden}) to extract massive young
objects from the MSX PSC. We will therefore use the criterion
F$_E$/F$_D$~$\geq$~2 to confirm the young nature of sources
selected on the basis of their red ISOGAL colours.

%%%%%

%%%
% \section{Near-infrared spectroscopic classification}
%%%

\subsection{\label{sec-sofi}Near-infrared spectroscopic observations}

We have conducted near-infrared (H and K bands) spectroscopic observations
with the SOFI instrument on the NTT in July 2000;
the observations and their analysis have been extensively described by
Schultheis et al. (\cite{ref-sofi}).
The purpose of these
observations was to derive the nature of sources with various positions in
the [15] vs. [7]--[15] colour magnitude diagram from their spectra.
Indeed, near infrared spectra contain enough
features to distinguish late type evolved stars from young objects:
the former show strong molecular bands (mainly CO and H$_2$O),
while the latter generally show only atomic lines (hydrogen Br$\gamma$,
HeI, FeII, NaI and CaI, among others),
or red featureless continuum, possibly with signatures of shocks
(e.g.~H$_2$ at 2.12\mic), characteristic of massive YSOs
(Hoare et al.~\cite{ref-hoare}).
It is therefore possible to
infer the nature of the sources from the equivalent widths of the
observed molecular bands, as shown in Schultheis et al. (\cite{ref-sofi},
see in particular their Figure 5).

\begin{figure}[htbp]
\begin{center}
\resizebox{8.5cm}{!}{ \rotatebox{0}{\includegraphics{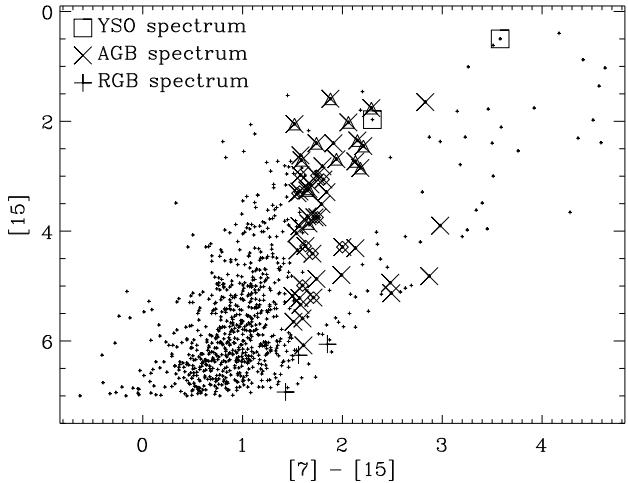}}}
\caption{\label{cmd_sofi}
Colour magnitude diagram for all sources in the FC--027 field, showing
the results of near-infrared spectroscopic classification. Sources with
spectra of YSO or hot stars
are shown with square symbols, while those with AGB or RGB spectra
are shown with $\times$ or large plus symbols, respectively.
Sources with AGB spectra also associated with LPV or OH/IR stars are
indicated with diamond or triangle symbols, respectively.}
\end{center}
\end{figure}

The observed sample included 65 sources in the FC--027 field.
Only two of them show near-IR spectra consistent with a young
stellar nature, or with hot stars in a more evolved stage.
These are 174516.2--290315 (labelled A33 in Schultheis et al.
\cite{ref-sofi}), also associated with a radio and an IRAS sources;
and 174504.9--291146 (labelled A40), which also has a radio counterpart.
The near-IR spectrum of the latter can possibly be interpreted
as tracing a Wolf-Rayet star. As can be seen in Fig.~\ref{cmd_sofi},
these observations tend to confirm that most sources with [7]--[15]
colours below or close to 2~mag are evolved stars in the AGB phase.

Five sources with colours [7]--[15]~$\ga$~2.5~mag show near-IR spectra
interpreted as AGB (Fig.~\ref{cmd_sofi}). In addition, the weak
([15]~=~4.87, [7]--[15]~=~1.74) source 174521.9--291344
has an AGB-type near-IR spectrum, but also has a radio continuum
counterpart (Fig.~\ref{cmd_yso}). A visual inspection of the images
shows that, for these six sources, either a second source or an
extended patch of diffuse emission shows up only at 15\mic.
Thus, these are certainly blends between AGB stars, mainly
detected in K-band and at 7\mic, and unrelated diffuse emission
that does not appear at wavelengths shorter than 15\mic, and
could possibly trace recent star formation.

\subsection{\label{sec-summary1}Summary}

\begin{figure*}[htbp]
\begin{center}
\resizebox{18cm}{!}{\includegraphics{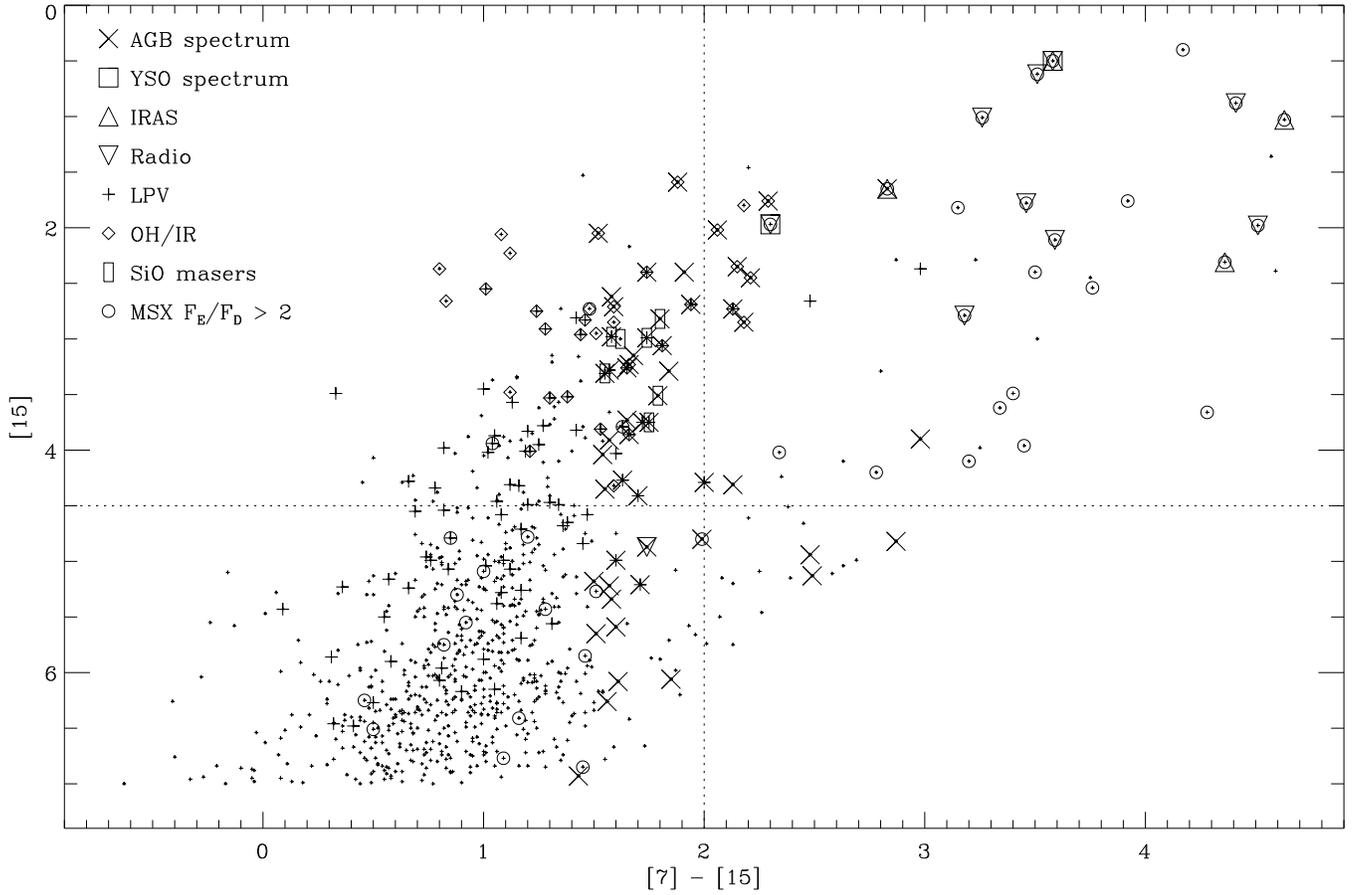}}
\caption{\label{cmd_final}{
Summary colour magnitude diagram for all sources
detected at 7 and 15\mic ~in the FC--027 field. The symbols show the
positions in this diagram of sources identified with
known objects (LPVs, OH/IR, IRAS and radio \hii ~regions), and those
with near-infrared spectral classification. Sources with MSX associations
with flux ratios F$_{\rm E}$/F$_{\rm D}$~$>$~2 are indicated
with open circles, also when detected in E but not in D. The dotted
lines show the limits [7]--[15]~$=$~2 and [15]~=~4.5 (see text).
}}
\end{center}
\end{figure*}

The results of all identifications discussed in the previous
sections are shown in the
summary [15] vs. [7]--[15] colour magnitude diagram
(Fig.~\ref{cmd_final}). It can be seen that sources interpreted
as young objects (i.e. with IRAS \hii ~regions or
radio continuum counterparts, and the two sources with
NIR-spectra of hot stars), occupy a well-defined
region in the right-upper part of this diagram, while sources identified
with AGB stars are mostly found to the left of [7]--[15]~$=$~2.
Only seven OH/IR stars are located in the [7]--[15]~$>$~2,
[15]~$<$~4.5 region. Six of them also have MSX counterparts, all with
flux ratios F$_{\rm E}$/F$_{\rm D}$ less than 2, while all
identified young objects have MSX flux ratios F$_{\rm E}$/F$_{\rm D} \,
>$~2. Therefore, F$_{\rm E}$/F$_{\rm D} \, >$~2 seems to be
a good criterion to confirm the young nature of
YSO candidates.

Also note that the majority of sources brighter than [15]~$=$~4.5 have
known counterparts in either of the young object or AGB classes.
Indeed, in the area covered by Glass et al. (\cite{GLASS2})
observations, the nature of 69 out of 88 sources (almost 80\%)
with [15]~$<$~4.5
and a 7-15\mic ~association can be derived from existing data,
or from their red MSX D--E colours for five of them.
The fraction of unidentified sources becomes much larger beyond
this magnitude limit, where the distinction between young and
evolved objects also becomes less clear.

%\section{\label{sec-summary2}Selection criteria for YSO candidates}
\section{\label{sec-summary2}Extraction of the young stellar population}

\subsection{\label{sec-sel-crit}Selection criteria}

According to the summary colour magnitude diagram (Fig.~\ref{cmd_final}
and Sect.~\ref{sec-summary1}), the simple colour criterion [7]--[15]~$>$~2
seems appropriate to select young objects, especially
for the brightest objects ([15]~$<$~4.5~mag). However, a moderate number
of evolved stars (e.g.~deeply embedded OH/IR stars)
also show [7]--[15] colours between 2 and 3~mag. Carbon stars
could also have [7]--[15] colours in excess of 2~mag, but it is known
that such objects are very rare in the direction of the inner \GB
~(e.g. Blanco et al.~\cite{ref-blanco}, Glass~\cite{ref-glass86}).
Consistently, we found no carbon stars among the $\sim$100 sources
in the inner \GB ~for which we obtained near-IR spectra (Schultheis et
al.~\cite{ref-sofi}). Planetary nebulae may also show red mid-IR
colours, but Jacoby \& Van de Steene (\cite{ref-PNsurvey}) estimated
their total number in the central 16~deg$^2$ of the Galaxy to be
of order 250, so that we expect a very small number of such objects
in this $\la$0.1~deg$^2$ field.

Another important parameter of the ISOGAL sources, related to their spatial
extension, appears to give a powerful criterion to distinguish massive star
forming regions from stars in their late evolution. The uncertainty in the
measured magnitude, especially at 15\mic, noted $\sigma_{15}$ in the
ISOGAL PSC (Schuller et al. \cite{PSC_FS}), is derived from the residuals
between the source profile and the analytical PSF used for the extraction.
While the values of $\sigma_{15}$ show a larger dispersion for faint sources,
and can reach typically 0.1--0.15~mag, they are generally low
($\leq$0.05~mag) for point sources brighter than [15]$=$4.5~mag.

\begin{figure}[htbp]
\begin{center}
\resizebox{8.5cm}{!}{ \includegraphics{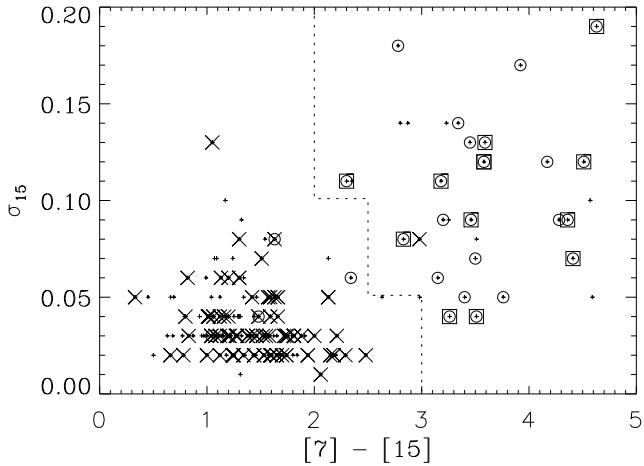}}
\caption{\label{sigma_coul_diag}$\sigma_{15}$ vs. [7]--[15] diagram
for all sources detected at both wavelengths and brighter than [15]$=$4.5
in FC--027. The sources associated with AGB stars are shown with cross
symbols, and those associated with young objects (IRAS \hii ~regions
or radio sources)
are indicated with square symbols. Sources with MSX associations
with F$_{\rm E}$/F$_{\rm D}$~$>$~2 are indicated with open circles.
The dotted line shows the criteria that we have used to extract
candidate young objects from the ISOGAL catalogue (Sect.~\ref{sec-sel-crit})}
\end{center}
\end{figure}

The distribution of the $\sigma_{15}$ values is shown versus the [7]--[15]
colours for all sources with [15]~$<$~4.5 in Fig.~\ref{sigma_coul_diag}.
Only one LPV star is found with [7]--[15]~$>$~2 and $\sigma_{15}$~$>$~0.05,
but its mid-infrared
magnitudes are affected by a blend with a bright nearby extended source.
On the other hand, all sources interpreted as young objects have high values
of both [7]--[15] and $\sigma_{15}$.
In other terms, stars in the AGB phase appear point-like
in the ISO observations, while more extended emission is detected
around young sources: these may be small groups or clusters of
YSOs, or single objects embedded in diffuse emission.
Indeed, ground-based mid-IR images of similar
objects selected from the MSX survey show, in most cases, a point-like
source embedded in extended diffuse emission, probably arising
from PAH (Hoare et al.~\cite{ref-hoare}).

Thus, we can combine
a colour and an extension criteria in order to extract the population
of ISOGAL sources associated with young objects, with
very little contamination by evolved stars. As shown in
Fig.~\ref{sigma_coul_diag}, the criteria that we have defined are:
\begin{equation}
\label{eq-criteria}
\left\{  \begin{array}{l}
2 \, \leq \, [7]-[15] \, < \, 2.5 \, \mbox{~and~} \, \sigma_{15} \, > \, 0.1 \\
2.5 \, \leq \, [7]-[15] \, < \, 3 \, \mbox{~and~} \, \sigma_{15} \, > \, 0.05 \\
\lbrack7]-[15] \, \geq \, 3, \, \mbox{no constraint on~} \, \sigma_{15} \\
\end{array} \right.
\end{equation}

Most sources with an MSX association with a flux ratio
F$_{\rm E}$/F$_{\rm D}$~$>$~2 are found in the same region of this
diagram, as shown with open circles in Fig.~\ref{sigma_coul_diag}.
This gives an additional argument showing that the combination of
the proposed colour and spatial extension criteria are appropriate
to select candidate young objects.

As compared with the selection criteria that we defined in Felli et al.
(\cite{Felli2002}), the addition of a constraint on the spatial extension
at 15\mic ~for the sources with moderately red colours
(2~$\leq$~[7]--[15]~$<$~3) greatly reduces the contamination of our sample
by anomalously red AGB stars (e.g. OH/IR stars with large
mass loss, Fig.~\ref{sigma_coul_diag}).
Also, we consider only sources redder than [7]--[15]~$=$~2 as
candidate young objects, while Felli et al. (\cite{Felli2002}) used
[7]--[15]~$\geq$~1.8 to select all candidates from the ISOGAL database.
However, both the interstellar extinction and confusion issues due to a high
source density are more extreme in the FC--027 field, located near
the Galactic Centre, than in the general Galactic Disk environment of
most ISOGAL fields. In addition, OH/IR and similar evolved
stars are relatively rare outside of the \GB: in the $\vert b \vert
< 2\dd$ range, the number density of IRAS sources with a probability
of variability greater than 90\% (i.e.~mostly AGB stars) already drops
by a factor $\sim$2 from $l \sim 0\dd$ to $\vert l \vert \sim 20\dd$,
while the number density of sources detected at 25, 60 and 100\mic
~(mostly young stars) is roughly constant over $\vert l \vert < 80\dd$
(see Fig. VII.Ap.7 and VII.Ap.16 in the IRAS Explanatory Supplement).
Therefore, we do not expect many evolved sources
in the Felli et al. (\cite{Felli2002}) sample extracted with
somewhat looser criteria. On the other hand, we may have missed a few
young objects with [7]--[15] colours below 2~mag in the present study.

We also want to stress that our criterion on $\sigma_{15}$ selects
sources marginally resolved with ISO, i.e. with angular sizes of
order 30$''$. This roughly corresponds to a 1~pc linear size at the
distance of the \GC. Thus, our combination of colour, magnitude
and spatial extension criteria is well adapted to select young
objects in the direction of the \GB, where most sources are expected
to be at similar distances (see also Sect.~\ref{sec-distance} below),
but it would be much less efficient in sight lines where distances to
sources are spread over a wide range.

%% Section generalisation to the ISOGAL-only sources
\subsection{\label{sec-extract-all}Extraction of all YSO candidates}

Using the proposed selection criteria (see Eq.~\ref{eq-criteria}),
33 candidate young objects with [15]~$<$~4.5~mag
(or F$_{15} \, > \,$280~mJy)
were selected, including 13 with IRAS and/or radio continuum
counterparts.
%This also includes the source 174449.5--291805, for which
%the 7--15\mic ~association was missed (Sect.~\ref{sec-iso-iras}).
%One of them, 174456.1--290604 is associated with the LPV star 22-136
%already discussed above, and has to be removed from this sample.
In addition, 15 sources are found with
[15]~$<$~4.5~mag, but with no association at 7\mic ~in the ISOGAL
PSC (which has a 50\%-completeness level around 8.0~mag at 7\mic,
see Sect.~\ref{sec-iso-obs}). One of them (174529.1--290413, with
[15]~=~4.41 and $\sigma_{15}$~=~0.13) is associated with an LPV
star; the lack of magnitude at 7\mic ~and the high $\sigma_{15}$
value can be explained by
confusion issues. The other 14 sources with [15]~$<$~4.5 all
have $\sigma_{15}$~$\geq$~0.07, and can also be considered as
candidate young objects.

\begin{figure}[htbp]
\begin{center}
\resizebox{8.5cm}{!}{ \includegraphics{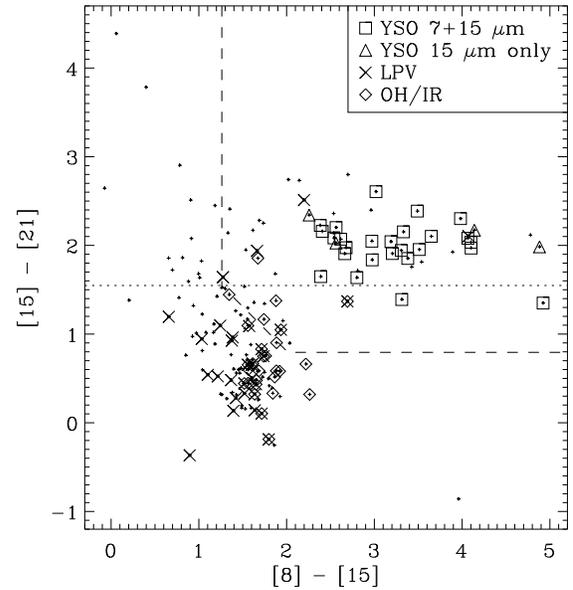}}
\caption{\label{fig-ccd-msx}MSX colour colour diagram for all sources
associated with ISOGAL in FC--027. The YSO candidates that we selected
from this field are indicated with square and triangle symbols. Sources
associated with LPV or OH/IR stars are shown with crosses and diamonds.
The dotted horizontal line corresponds to F$_{\rm E}$/F$_{\rm D}$~$=$~2,
and the dashed lines show the criteria used by Lumsden et al.
(\cite{ref-lumsden})}
\end{center}
\end{figure}

Among these 46 candidates, 30 (or 65\%) are also associated
with MSX sources. All of them are detected in the A, D and E bands,
and all but two have flux ratios F$_{\rm E}$/F$_{\rm D}$~$>$~2,
as can be seen in Fig.~\ref{fig-ccd-msx}. On the other hand, only
four LPV stars and one OH/IR star have F$_{\rm E}$/F$_{\rm D}$
above 2. This colour-colour diagram therefore seems to be an
efficient means for distinguishing young objects from evolved stars,
whereas the mid-IR selection criteria used by Lumsden et al.
(\cite{ref-lumsden}) do not reject evolved stars as efficiently,
as can be seen in Fig.~\ref{fig-ccd-msx} - however, they used
additional criteria based on 2MASS near-IR magnitudes to eliminate
many of the evolved sources.

The most relevant infrared properties of our 46 YSO candidates
are given in Tables~\ref{tab-list1} and \ref{tab-list2} in
Appendix~\ref{sec-append}. Tables with all available magnitudes
(ISO, DENIS, 2MASS, MSX and IRAS when relevant) can be retrieved
electronically through CDS.
These tables also contain aperture magnitudes at 15\mic. These
were computed by integrating a number of pixels in an inner radius,
and subtracting the background estimated as the median in an
outer annulus; the radii that we used were adapted for every
individual source, in order to avoid contaminations by nearby
sources.

%\subsection{Comments on individual sources}
%
%\subsection{\label{sec-cluster}Comments on the cluster of sources
\subsection{\label{sec-cluster}Comments on the cluster
around LS95~7}

\begin{figure}[htbp]
\begin{center}
\resizebox{8.5cm}{!}{ \rotatebox{0}{\includegraphics{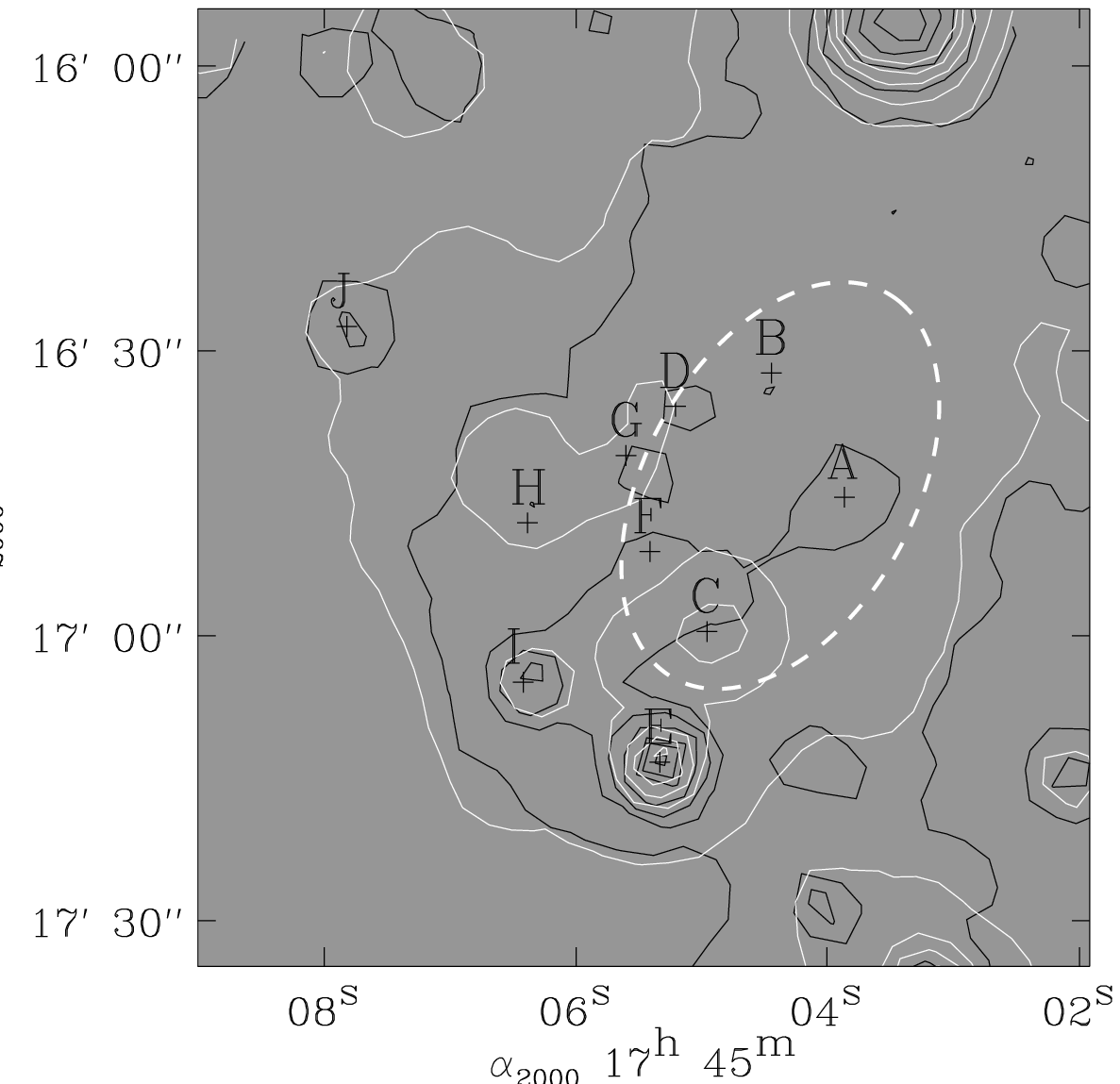}}}
\caption{\label{im_cluster}Contour plots of the ISOCAM LW5 (7\mic, black contours)
and LW9 (15\mic, white contours) around ($\alpha$,$\delta$) = ($17^h45^m05^s$,
$-29\dd16'50''$). The LW5 contours correspond to flux densities of 40, 60,
80, 110, 150, 200 and 300 mJy/pixel, and the LW9 contours correspond to
60, 100, 140, 180, 220 and 260 mJy/pixel. The crosses and letters indicate
the positions of extracted ISOGAL sources, and details are given in
Table~\ref{tab-cluster}. The ellipse drawn with thick dashed line shows
the position and size (FWHM) of the radio source LS95~7. The very bright
source at the limit of the field to the north is associated with
LS95~8.}
\end{center}
\end{figure}

A compact group of nine ISOGAL point sources (Fig.~\ref{im_cluster})
is located in the vicinity of the radio source No.~7 detected at
1.6~GHz by Liszt and Spiker (\cite{LS95}), who reported an
extended size of 31$''$$\times$48$''$ (Sect.~\ref{sec-radio}).
Sources extracted in a 30$''$ radius around 17$^h$45$^m$05$^s$,
$-29\dd16'50''$ are reported in Table~\ref{tab-cluster}.
At least two of them, sources C and H, are probably young
objects, since they
have [7]--[15] colours above 3.5~mag, and signs of spatial extension
as indicated by value of $\sigma_{15}$ above 0.1~mag.
These two sources are in the samples extracted with our criteria
discussed in Sect.~\ref{sec-extract-all}.

\begin{table}[htbp]
\caption{\label{tab-cluster}Coordinates, 7 and 15\mic ~magnitudes,
extension at 15\mic ~$\sigma_{15}$ and 7-15\mic ~association quality
flag for the sources located within the cluster. The Id. column gives
the identification letters, as they appear in Fig.~\ref{im_cluster}.}
\begin{center}
\begin{tabular}{llllll}
\hline
\hline
Id. & ISOGAL name & [7] & [15] & $\sigma_{15}$ & q7-15 \\
    &  ISOGAL-PJ  & mag & mag & mag & $\leq$4 \\
\hline
A & 174503.9--291645 &  6.59 &   -   &  -   & 0 \\
B & 174504.4--291632 &  7.63 &   -   &  -   & 0 \\
C & 174505.0--291659 &   -   &  3.21 & 0.11 & 0 \\
D & 174505.2--291635 &  7.39 &   -   &  -   & 0 \\
E & 174505.3--291713 &  4.21 &  2.73 & 0.04 & 4 \\
F & 174505.4--291651 &  6.80 &   -   &  -   & 0 \\
G & 174505.6--291641 &  7.48 &   -   &  -   & 0 \\
H & 174506.4--291648 &  7.41 &  3.96 & 0.13 & 2 \\
I & 174506.4--291704 &  5.35 &  3.89 & 0.05 & 4 \\
%J & 174507.8--291627 &  6.18 &  5.30 & 0.09 & 4 \\
\hline
\end{tabular}
\end{center}
\end{table}

Diffuse emission is also visible
around these sources at 15\mic, suggesting that this could be a
group of young objects still in their parent cocoon. We can also note
that if this group is at the distance of the \GC, the projected distance
between sources C and H would be of order 1~pc, which could correspond
to a very young OB-association. Complementary observations at higher
spatial resolution, with a sensitivity at least comparable to ISOGAL,
would greatly help in understanding the nature of this object.

\subsection{\label{sec-distance}Distance to the sources}

Most radio sources that we considered in the present paper were
only observed in the continuum, so that no radial velocity data
are available. Only for the source LS95~7 that we discussed in
the previous section, Liszt and Spiker (\cite{LS95}) report a
radial velocity V$_{\rm LSR}$~=~--44.1~\kms, as derived from
observations in the H70$\alpha$ radio recombination line. Thus,
this source is probably associated with the Sgr~C complex, in
the vicinity of the \GC.

Using the available data, we cannot exclude that some of the
candidates that we have selected are foreground. However,
given the high over-density of such sources in the inner \GB ~as
compared to the rest of the Galactic disk, we expect a very
small fraction of foreground sources: the Besan\c{c}on model
(Robin et al.~\cite{ref-besancon}) predicts $\sim$8\%
contamination by foreground objects.
It is still possible that a few of our candidates are much
closer than the \GC, especially among the brightest sources.
Additional observing programmes are ongoing to further constrain
the distance to these sources.

\subsection{Summary}

\begin{figure*}[htbp]
\begin{center}
\resizebox{15cm}{!}{ \rotatebox{0}{\includegraphics{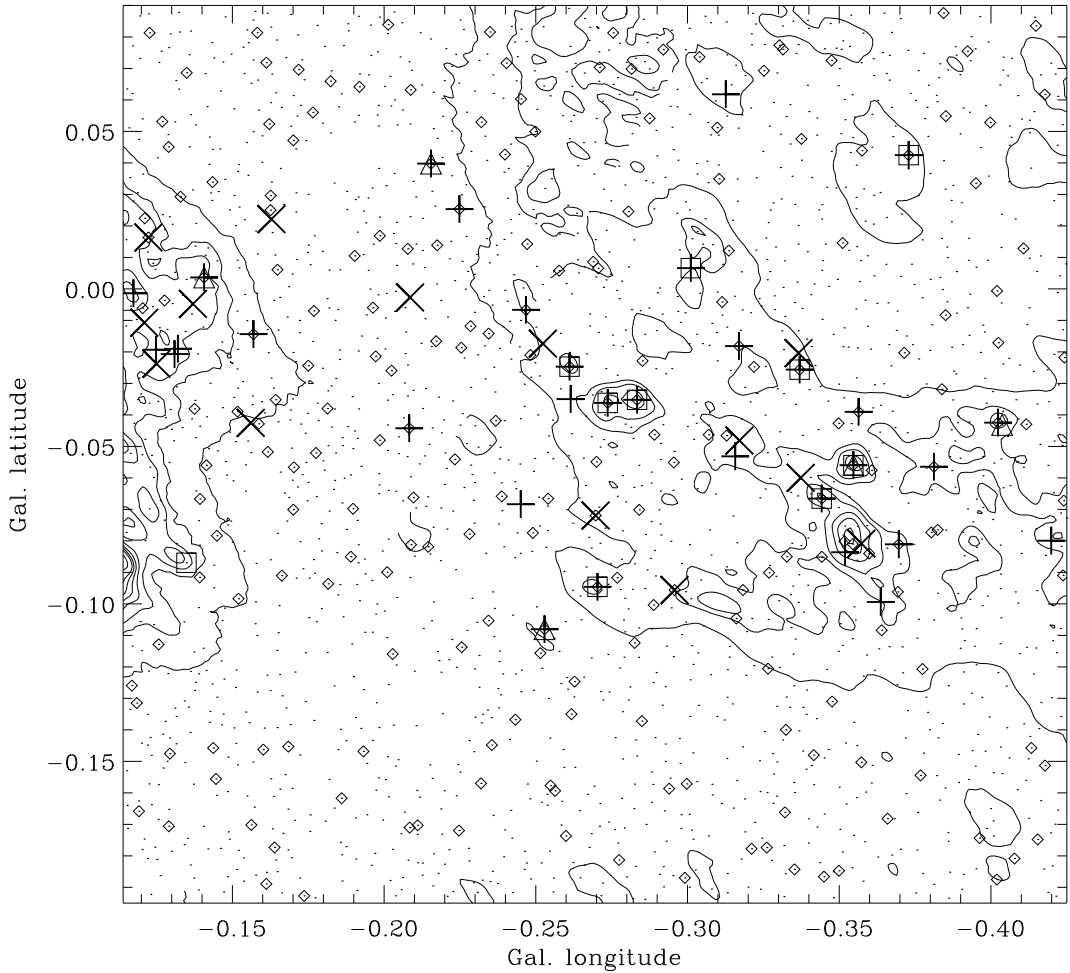}}}
\caption{\label{fig-mapYSO}Distribution in galactic coordinates of
our 46 candidates. All 2043 sources from the ISOGAL catalogue
are represented with small dots. The 32 candidates with 7 and
15\mic ~detections are shown with plus signs, and the 14 candidates
only seen at 15\mic ~with crosses. Sources with IRAS counterparts
are indicated by triangles and those with radio continuum by squares
(see also the IRAS uncertainty ellipses in
Fig.~\ref{fig-iras-ellips}).
The sources associated with MSX point sources are shown
with small diamonds. The contours show the radio continuum emission
at 1.6~GHz (data were kindly provided by Dr.~H.S.~Liszt), with levels
at 0., 0.01, 0.02, 0.04, 0.06, 0.08, 0.1, 0.15, 0.2 and 0.25 Jy/beam}
\end{center}
\end{figure*}

We have been able to extract a sample of 46 candidate massive
young objects brighter than 4.5~mag at 15\mic ~(or
F$_{\nu} \, > \,$280~mJy)
from a 0.09~deg$^2$ area at about 40~pc in projection from
the \GC, using mid-infrared photometry from
the ISOGAL survey. We have adopted a conservative approach, since
this survey goes down to magnitude 7.0 at 15\mic ~with good
completeness (Sect.~\ref{sec-iso-obs}); the distinction
between young and evolved stars becomes more subtle when
going to fainter levels than this. A source located at 8.5~kpc with
magnitude [15]~$=$~4.5 would have a luminosity around
10$^4$~\lsol ~according to Felli et al. (\cite[Sect.~3.2]{Felli2002}),
consistent with a single B0 star.

Only 13 of these 46 sources show other signatures of young
massive stars, such as IRAS or radio continuum counterparts (including
one not seen by MSX), and 16 do not appear in the MSX PSC,
while most of the others display MSX colours typical
of young stars.
The distribution of these YSO candidates is shown in galactic
coordinates in Fig.~\ref{fig-mapYSO}. It appears that most of
them are located in the $-0.1\dd \leq b
\leq 0\dd$ range, and that their distribution is tilted
with respect to the Galactic Plane, going from $b \sim
0\dd$ at $l \sim -0.1\dd$ to $b \sim -0.1\dd$
at $l \sim -0.4\dd$. In the mid-infrared, diffuse emission
probably arising from PAHs and small dust grains is seen in
the same latitude range (Fig.~\ref{images}).
Moreover, strong radio continuum emission
at 1.6~GHz is observed at the positions of most candidates.
Some of them are coincident with peaks of radio emission, and
others are located in regions of extended radio emission, including
eight candidates without known counterparts located in the extended
component around Sgr~A. This adds to the confidence that
we are interpreting these sources correctly as young massive
objects.

\section{Conclusion}

We have shown in this paper that the bright end of the stellar
populations detected by ISOGAL can be well characterised, using
already existing data and complementary mid-infrared observations.
A sample containing 182 sources brighter than 280~mJy at 15\mic,
complete over a 0.09~deg$^2$ area close to the \GC,
has been associated with several catalogues of evolved stars
on one hand, and with high-mass young objects previously
detected by IRAS or in the radio continuum on the other.
With the addition of infrared photometry at 15 and 21\mic
~from MSX, and spectroscopic near-infrared data for some
sources, a majority (61\%) could be identified
with evolved stars or with young objects.

The mid-infrared magnitudes and colours and the spatial
extension of the 15\mic ~emission of the young objects
are sufficiently peculiar to provide robust criteria
for selecting good candidate young massive objects from
the ISOGAL data. Using a colour criterion [7]--[15]~$\geq$~2~mag,
combined with a minimum spatial extension indicated by
$\sigma_{15}$~$>$~0.1~mag for the least red sources ([7]--[15]
between 2 and 2.5~mag), we could extract 46 candidates from
the analysed field. Only 13 of them were previously
reported as ultra-compact \hii ~regions. About 65\%
are also detected by MSX, and they also have red
colours in the longer 15 to 21\mic ~wavelengths. Their positions
in the field are almost always coincident with diffuse extended
infrared emission, with peaks or more extended emission in
the radio continuum in most cases. They are mainly distributed
in the $-0.1\dd \, \leq \, b \, \leq \, 0\dd$ range,
or very close to the Galactic Plane, where one expects
to find the largest incidence of star formation.
All this tends to confirm their interpretation as young
objects, embedded in vast structures of dust and gas heated by
recently formed stars.

The exact nature of our objects remains unclear. Following Felli
et al. (\cite{Felli2002}), their bolometric luminosities can be
roughly derived from their flux densities at 15\mic, and one finds
values in the range $10^4-10^6$~\lsol, assuming that they are
at the distance of the \GC. Such luminosities could come from
single O or early B stars. However, most of them show spatial
extension of order 1~pc (as resolved by ISO) and only very
few are associated with known ultra-compact \hii ~regions,
suggesting that there is no very early type star present
in most cases. Thus, these objects could be small groups or
clusters of young stars of intermediate masses. To be detected
with such red colours in the mid-infrared, they would have to
be in a protostellar phase, possibly in the compact
molecular core stage, prior to the formation of a compact
\hii ~region (e.g.~Brand et al.~\cite{ref-brand}),
or an evolutionary stage somehow similar to class I for
low-mass YSOs.
Some candidates may also be compact \hii ~regions
falling below the sensitivity of current radio continuum
surveys, thus corresponding to the next stage in the early
evolution of massive stars.

Several follow-up projects, including spectro-imagery with
the Spitzer Space Telescope, have been initiated to better
understand the nature of these objects. Moreover, the ISOGAL data
cover most parts of the inner Galactic bulge, in the $\pm 1.5\dd$
longitude range. The criteria that we have presented
here can be used to extract the young population from
the other ISOGAL fields, to get a more complete picture of
the recent star formation in the central $\sim$400~pc of
the Galaxy; this will be the subject of a second paper.

\begin{acknowledgements}
We are very grateful to Dr.~H.S.~Liszt for providing us with his
VLA 1.6~GHz data of the Galactic Centre region. 
This research made use of data products from the Midcourse Space
Experiment. Processing of the data was funded by the Ballistic
Missile Defense Organization with additional support from NASA
Office of Space Science.
We gratefully thank the MSX team and in particular
Dr.~D.~Mizuno for making us parts of the MSX catalogue
version 2.3 available prior to its publication.
MS is supported by the APART programme of the Austrian Academy
of Science.

\end{acknowledgements}

% ---------------------- APPENDIX ------------------------------
\appendix
\section{Complete list of YSO candidates}
\label{sec-append}

Table \ref{tab-cds} shows two examples of the full data for
our 46 candidate young objects. This table contains photometry
and quality flags from the 2MASS, DENIS, ISOGAL, MSX and
IRAS catalogues. The separations between the ISOGAL objects
and the other sources are also given, as well as the
separations between the 7\mic ~and 15\mic ~sources.
The ISOGAL 15\mic ~flux densities have been determined by
PSF-fitting (column F$_{15}$) and by aperture photometry
(column ap$_{15}$).
The complete version of this table is available through
VizieR at the CDS.

Tables \ref{tab-list1} and \ref{tab-list2} show only the
ISOGAL and DENIS data for the candidates detected at
7 and 15\mic, and the ones only detected at 15\mic,
respectively. These tables are available in the online
version of the present paper.

\begin{table*}[htbp]
\begin{minipage}{\linewidth}
\caption{\label{tab-cds}2MASS and DENIS magnitudes, ISOGAL, MSX and IRAS
fluxes for two among our 46 candidate young objects. The full table
is available through Vizier at the CDS$^1$.}
\begin{tabular}{l|lllll|llll}
\hline
\hline
 & \multicolumn{5}{|c}{2MASS} & \multicolumn{4}{|c}{DENIS} \\
Name & J & H & K & flag & sep. & I & J & K & sep. \\
ISOGAL-PJ- & [mag] & [mag] & [mag] & & [$''$] &
[mag] & [mag] & [mag] & [$''$] \\
\hline
174437.9--291014 & 14.640 & 12.895 & 11.750 & BUU & 1.62 &
- & - & - & - \\
174449.5--291805 & 16.116 & 13.162 & 11.189 & UBA & 0.61 &
- & 15.31 &  9.02 & 1.49 \\
\hline
\end{tabular}

\vspace{2mm}

\begin{tabular}{llllll|lllllll}
\hline
\hline
\multicolumn{6}{c}{ISOGAL} & \multicolumn{7}{|c}{MSX} \\
F$_7$ & F$_{15}$ & ap$_{15}$ & $\sigma_{15}$ & flag & sep. &
Name & F$_{\rm A}$ & F$_{\rm C}$ & F$_{\rm D}$ & F$_{\rm E}$ & 
flag & sep. \\
$\lbrack$Jy] & [Jy] & [Jy] & [Jy] & & [$''$] &
 & [Jy] & [Jy] & [Jy] & [Jy] & & [$''$] \\
\hline
 0.132 &  0.370 &  4.66 & 0.067 & 33 & 4.68 &
G359.6885+00.0634 & 0.49 & 0.96 & 2.49 & 13.20 & 4444 & 7.1 \\
0.270 &  1.854 &  3.70 & 0.255 & 43 & 6.00 &
G359.5969--00.0415 & 0.66 & 2.02 & 4.43 & 15.47 & 4444 & 3.7 \\
\hline
\end{tabular}

\vspace{2mm}

\begin{tabular}{llllll}
\hline
\hline
\multicolumn{6}{c}{IRAS} \\
Name & F$_{12}$ & F$_{25}$ & F$_{60}$ & F$_{100}$ & flag \\
 & [Jy] & [Jy] & [Jy] & [Jy] & \\
\hline
- & - & - & - & - & - \\
17416-2916 & 12.9 &  43 & 1480 & 4850 & 1311 \\
\hline
\end{tabular}
\end{minipage}

\vspace{1mm}
$^1$\url{http://vizier.u-strasbg.fr/viz-bin/VizieR}
\end{table*}

\begin{table*}[htbp]
\caption{\label{tab-list1}Main characteristics of the 32 YSO candidates
detected at 7 and 15\mic: J and K DENIS magnitudes, ISO--DENIS 
association quality flag q$_{ID}$
and separation; magnitudes, uncertainties and quality flags at 7 and
15\mic, 7--15\mic ~association quality flag q$_{II}$ and separation;
aperture magnitude at 15\mic; MSX: this column contains a `+' if
the source is associated (within 8$''$) with an MSX source with
F$_{\rm E}$/F$_{\rm D}$~$>$~2, and a `X' if there is an association,
but with F$_{\rm E}$/F$_{\rm D}$~$<$~2.}
\begin{center}
\begin{tabular}{lllllllllllllll}
\hline
\hline
ISOGAL name     &   J   &   K   & q$_{ID}$ & sep. & [7] & $\sigma_7$ & q$_7$ &
[15] & $\sigma_{15}$ & q$_{15}$ & q$_{II}$ & sep. & ap$_{15}$ & MSX \\
ISOGAL-PJ       & mag   &  mag  & $\leq5$ & $''$ & mag & mag & $\leq4$ &
mag  & mag & $\leq4$ & $\leq4$ & $''$ & mag & \\
\hline
174433.8-291355 & --    & --    & -- & --  & 6.49 & 0.10 & 4 & 
 1.98 & 0.12 & 4 & 4 & 1.30 & 1.10 & + \\
174437.9-291014 & --    & --    & -- & --  & 6.98 & 0.16 & 3 & 
 4.20 & 0.18 & 3 & 2 & 4.68 & 1.45 & + \\
174449.5-291805 & 15.31 &  9.02 & 4 & 1.49 & 6.20 & 0.12 & 4 & 
 2.45 & 0.14 & 3 & 0 & 6.59 & 1.70 & + \\
174452.5-291122 & 11.39 &  7.66 & 2 & 3.05 & 5.70 & 0.18 & 3 & 
 2.11 & 0.13 & 4 & 2 & 3.95 & 0.60 & + \\
174454.9-291413 & --    & --    & -- & --  & 4.27 & 0.09 & 4 & 
 1.01 & 0.04 & 4 & 4 & 0.91 & 0.55 & + \\
174455.3-291538 & --    & --    & -- & --  & 4.97 & 0.12 & 4 & 
 1.82 & 0.06 & 4 & 4 & 0.46 & 1.40 & + \\
174455.8-292009 & --    & 10.51 & 4 & 1.76 & 7.23 & 0.13 & 4 & 
 3.98 & 0.09 & 4 & 3 & 2.12 & 2.80 &  \\
174455.8-291727 & 11.92 & 10.62 & 2 & 3.46 & 6.98 & 0.09 & 4 & 
 2.39 & 0.05 & 4 & 4 & 0.42 & 1.90 & X \\
174456.1-291257 & --    & --    & -- & --  & 5.93 & 0.05 & 4 & 
 1.36 & 0.10 & 4 & 4 & 0.41 & 0.60 & X \\
174457.0-290557 & 15.25 & 10.11 & 5 & 0.63 & 4.48 & 0.17 & 3 & 
 1.65 & 0.08 & 4 & 4 & 0.94 & 0.70 & + \\
174459.1-290653 & --    & 11.07 & 4 & 1.47 & 6.89 & 0.10 & 4 & 
 3.49 & 0.05 & 4 & 4 & 0.76 & 3.00 & + \\
174459.5-291604 & --    & 10.78 & 3 & 1.81 & 4.08 & 0.17 & 3 & 
 0.50 & 0.12 & 3 & 4 & 1.33 & --0.30 & + \\
174503.2-291737 & --    & --    & -- & --  & 5.90 & 0.06 & 4 & 
 2.40 & 0.07 & 4 & 4 & 0.47 & 1.90 & + \\
174503.4-290900 & --    & --    & -- & --  & 5.68 & 0.18 & 3 & 
 1.76 & 0.17 & 3 & 4 & 1.82 & 0.00 & + \\
174503.5-291552 & --    & 11.06 & 2 & 2.96 & 4.13 & 0.10 & 4 & 
 0.62 & 0.04 & 4 & 4 & 1.67 & 0.35 & + \\
174504.4-291359 & --    & --    & -- & --  & 6.96 & 0.17 & 3 & 
 3.62 & 0.14 & 3 & 3 & 2.95 & 2.00 & + \\
174504.9-291146 & 15.24 &  9.01 & 5 & 1.11 & 4.27 & 0.09 & 4 & 
 1.97 & 0.11 & 4 & 4 & 1.66 & 0.70 & + \\
174505.6-291018 & --    & --    & -- & --  & 5.24 & 0.16 & 3 & 
 1.78 & 0.09 & 4 & 4 & 1.09 & 1.00 & + \\
174506.4-291648 & 13.51 &  9.17 & 5 & 0.34 & 7.41 & 0.12 & 4 & 
 3.96 & 0.13 & 3 & 2 & 3.54 & 1.50 & + \\
174506.5-291118 & --    & --    & -- & --  & 5.29 & 0.12 & 4 & 
 0.88 & 0.07 & 4 & 4 & 0.33 & 0.10 & + \\
174508.0-291039 & 14.44 &  9.03 & 5 & 0.67 & 6.59 & 0.09 & 4 & 
 4.24 & 0.11 & 4 & 2 & 4.38 & 2.80 &  \\
174508.4-291753 & --    & --    & -- & --  & 7.94 & 0.10 & 4 & 
 3.66 & 0.09 & 4 & 4 & 1.78 & 2.60 & + \\
174516.2-290315 & 11.47 &  7.86 & 5 & 1.05 & 5.66 & 0.15 & 3 & 
 1.03 & 0.19 & 3 & 2 & 3.42 & --1.00 & + \\
174517.8-290813 & --    & --    & -- & --  & 7.30 & 0.09 & 4 & 
 4.10 & 0.09 & 4 & 4 & 0.42 & 3.50 & + \\
174518.1-290439 & --    & --    & -- & --  & 6.30 & 0.10 & 4 & 
 2.54 & 0.05 & 4 & 4 & 1.66 & 1.95 & + \\
174518.1-291051 & --    & --    & -- & --  & 6.51 & 0.16 & 3 & 
 3.00 & 0.08 & 4 & 4 & 0.68 & 2.60 &  \\
174520.7-291258 & --    & --    & -- & --  & 5.97 & 0.19 & 3 & 
 2.79 & 0.11 & 4 & 4 & 1.55 & 1.70 & + \\
174520.7-290213 & --    & 10.40 & 2 & 2.89 & 4.57 & 0.18 & 2 & 
 0.40 & 0.12 & 3 & 4 & 0.61 & --0.40 & + \\
174522.8-290331 & --    & --    & -- & --  & 5.16 & 0.14 & 3 & 
 2.29 & 0.14 & 3 & 2 & 5.51 & 1.30$^{(\dagger)}$ &  \\
174523.3-290331 & 15.76 &  8.48 & 5 & 0.24 & 5.52 & 0.09 & 1 & 
 2.29 & 0.14 & 3 & 3 & 2.67 & 1.30$^{(\dagger)}$ &  \\
174523.9-290310 & 14.58 &  8.72 & 5 & 0.36 & 6.09 & 0.06 & 4 & 
 3.29 & 0.14 & 3 & 2 & 3.95 & 2.30 &  \\
174526.3-291229 & --    & --    & -- & --  & 6.67 & 0.16 & 2 & 
 2.31 & 0.09 & 4 & 4 & 0.46 & 1.60 & + \\

\hline
\end{tabular}
\end{center}
{\it Note}: $^{(\dagger)}$ These two 7\mic ~sources correspond to a
single source at 15\mic.
\end{table*}

\begin{table*}[htbp]
\caption{\label{tab-list2}Main characteristics of the 14 YSO candidates
detected only at 15\mic. Columns are the same as in Table~\ref{tab-list1},
except for the 7\mic ~and 7-15\mic ~association data.}
\begin{center}
\begin{tabular}{llllllllll}
\hline
\hline
ISOGAL name     &   J   &   K   & q & sep. &
[15] & $\sigma_{15}$ & q & ap$_{15}$ & MSX \\
ISOGAL-PJ       & mag   &  mag  & $\leq5$ & $''$ &
mag  & mag & $\leq4$ & mag & \\
\hline
174453.8-291402 & --    & --    & -- & --  & 3.79 & 0.09 & 4 & 3.30 &  \\
174502.9-291518 & --    & --    & -- & --  & 3.85 & 0.07 & 4 & 3.25 &  \\
174503.1-291354 & --    & --    & -- & --  & 3.78 & 0.09 & 4 & 3.40 &  \\
174505.0-291659 & --    & --    & -- & --  & 3.21 & 0.11 & 3 & 2.50 &  \\
174505.1-290937 & --    & --    & -- & --  & 3.83 & 0.11 & 4 & 3.25 &  \\
174508.0-290655 & --    & --    & -- & --  & 4.34 & 0.14 & 3 & 3.60 &  \\
174508.7-290348 & --    & 11.02 & 2 & 2.89 & 4.17 & 0.12 & 4 & 2.95 &  \\
174515.4-291213 & --    & --    & -- & --  & 3.40 & 0.11 & 4 & 2.70 & + \\
174515.9-290155 & --    & 11.12 & 3 & 2.12 & 2.63 & 0.13 & 3 & 1.95 & + \\
174517.3-291418 & --    & --    & -- & --  & 3.34 & 0.09 & 4 & 3.00 & + \\
174518.7-290320 & --    & 11.25 & 3 & 2.27 & 4.06 & 0.18 & 2 & 1.90 &  \\
174522.4-290242 & 14.88 &  9.15 & 3 & 2.25 & 3.68 & 0.17 & 3 & 1.50 &  \\
174524.8-290529 & --    & --    & -- & --  & 3.89 & 0.19 & 3 & 2.00 & + \\
174524.9-290318 & --    & --    & -- & --  & 3.33 & 0.14 & 3 & 1.90 &  \\

\hline
\end{tabular}
\end{center}
\end{table*}

% ------------------- END OF APPENDIX -------------------------

\end{document}